\documentclass{kluwer}    
\usepackage{graphicx}

\begin{document}
\begin{article}
\begin{opening}
\title{Tomographic simulations of accretion disks in Cataclysmic Variables - flickering and wind} 
\author{Fab\'{\i}ola Mariana Aguiar \surname{Ribeiro}}  
\author{Marcos \surname{Diaz}}
\runningauthor{Fab\'{\i}ola Ribeiro and Marcos Diaz}
\runningtitle{Tomographic simulations of accretion disks in CVs - flickering and wind}
\institute{Universidade de S\~ao Paulo - Instituto de Astronomia, Geof\'{\i}sica e Ci\^encias Atmosf\'ericas - Brazil \email{fabiola@astro.iag.usp.br}}
\date{Received: 2005 September 30 / Accepted: 2005 October 30}


\begin{abstract}
Cataclysmic Variables (CVs) are close binary systems where mass is transferred from a red dwarf star to a white dwarf star via an accretion disk. The flickering is observed as stochastic variations in the emitted radiation both in the continuum and in the emission line profiles.\\
The main goal of our simulations is to compare synthetic Doppler maps with observed ones, aiming to constrain the flickering properties and wind parameters.\\
A code was developed which generates synthetic emission line profiles of a geometrically thin and optically thick accretion disk. The simulation allows us to include flares in a particular disk region. The emission line flares may be integrated over arbitrary ``exposure'' times, producing the synthetic line profiles. Flickering Doppler maps are created using such synthetic time series. The presence of a wind inside the Roche lobe was also implemented. Radiative transfer  effects in the lines where taken into account in order to reproduce the single peaked line profiles frequently seen in nova-like CVs.
\end{abstract}
\end{opening}

\section{Introduction}

The flickering, or rapid variability, is observed as stochastic fluctuations in the emitted radiation. The typical timescales range from seconds to tenths of minutes, and the amplitudes vary from cents of magnitude to more than one magnitude. The flickering is a defining characteristic of cataclysmic variables, being frequently used to classify an object as a CV. Flickering is also observed in other classes of objects, as in some symbiotic stars \cite{Miko90}. The first CV where flickering was observed is UX UMa \cite{Line49}. Since then, many photometric studies were made aiming to locate the flickering source region in many systems. \inlinecite{Diaz01} proposed a tomographic method to map the flickering source regions using line profile data. The flickering tomography was applied to V442 Oph system, and an isolated flickering source region could not be identified. The objective of our simulations is to generate flickering tomograms from synthetic line profiles and compare these tomograms with the observed maps, aiming to constrain some parameters of the flickering and locate its forming region.\\
The presence of winds in cataclysmic variables is noticed by strong wind driven lines in UV, i.e. C IV, and the occurrence of P-Cygni profiles. As some tomograms present emission at low velocities, we have implemented the presence of a optically thick wind, aiming to reproduce this behavior. The comparison of model tomograms with observations may help us to constrain the main wind parameters.\\
In this proceeding we present the main physical concepts and parameterizations contained in the simulations and some preliminary results that arise from them.

\section{Simulations}

The first part of the simulation is the calculation of the line profiles from a Keplerian accretion disk. The disk steady line emissivity is described as a radial power law. In this disk, regions of enhanced emission could be marked to represent the hot spot and/or the boundary layer. The flickering and wind components are simply added to the underlying disk emission.

\subsection{Flickering}

\begin{figure}[H]
\begin{center}
\includegraphics[width=5.5in]{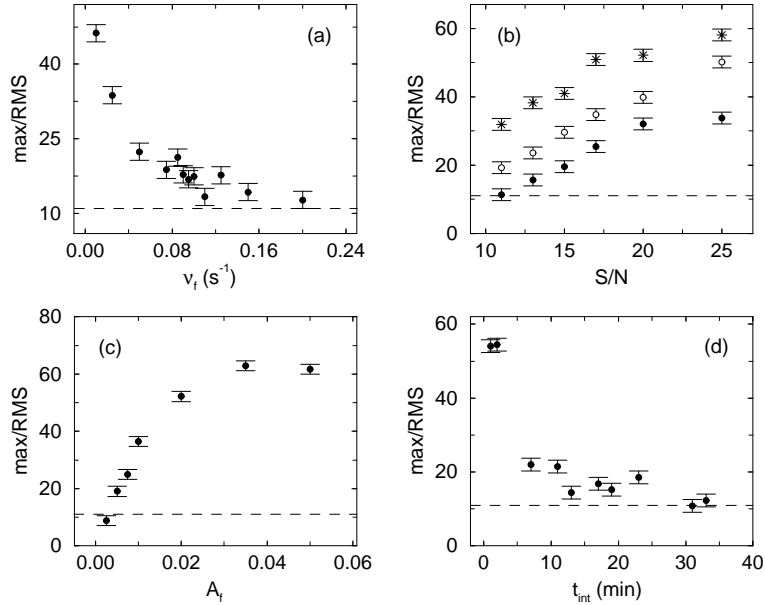}
\caption{Behavior of the quality factor (see text) with (a) flickering frequency, (b) S/N ratio, (c) flickering amplitude and (d) integration time. In (b) the ``$\bullet$'' correspond to simulations with 1000 spectra and flickering amplitude 0.5\%, the ``$\circ$'' were found with 500 spectra and amplitude 0.5\% of the total line flux, and the ``$\ast$'' correspond to simulations with 1000 spectra and amplitude of 1\%. The dashed lines indicate max/RMS equal 11.} \label{fig:1}
\end{center}
\end{figure}

The flickering is implemented as a set of random flares in a predefined region of the accretion disk. In the simulations presented here the flickering was set to occur inside the hot spot region. The flares have instantaneous rise and exponential decay. They are integrated over an ``exposure time'' to generate the flickering contribution to each synthetic spectra.\\
To quantify our ability of detecting flickering spots, a quality factor was defined as being the ratio between the spot intensity in the flickering tomogram and the RMS over a steady region. The simulations suggest a detection criterion corresponding to quanlity factors greater than 11. The quality factor behavior was verified by varying the flickering parameters one by one (fig. \ref{fig:1}). As expected, one verifies that the detection is better for high S/N data and a large number of spectra. We also verified that the greater the flickering amplitude, the better is the detection. The information contained in the low amplitude flickering is lost first by dilution in the noise. The detection is also better for low frequency flickering and short integration times. Furthermore, we verified that the information contained in the high frequency flickering is lost first. As expected, high frequency structures are lost when long integration times are used.

\subsection{Wind}

The wind was implemented as coming from the accretion disk with a bipolar geometry, limited by the white dwarf and the disk outer rim. The velocity field of the wind is a composition of a Keplerian component, due to the disk angular moment conservation, and a poloidal component. The poloidal component follows a Castor and Abbot velocity field \cite{Long02}. The Keplerian wind component is dominant over the poloidal one at small radii while both components are comparable at the outer disk.\\
Radiative recombination is considered as the main Balmer and He line production mechanism in the wind. Radiative transfer in the wind must be included to reproduce the observed Doppler tomograms filled at low velocities. Was found that an optical depth $\tau_{line} > 10$ is needed to reproduce single peaked line profiles. Scattering is not included in the radiative transfer. The emissivity is calculated for each wind element, then the correspondent absorption inside the primary Roche lobe and along the line of sight is taken into account.\\
The response of the line profiles to each wind parameter was checked out. For instance, lower values of the wind acceleration coefficient produce a deficit in the red line wing, which can be interpreted as the effect of the poloidal wind component at the opposite side of the Roche lobe being more attenuated due to self absorption. As the wind get less collimated, the line profiles become wider. Other parameters, like the terminal velocity, effective acceleration scale and wind initial velocity, basically change the total line intensity.

\section{Conclusions}

Simulations of synthetic accretion disc line profiles including flickering and wind were performed. From the simulations one can see that high S/N and high time-resolution spectra are needed in order to obtain information from flickering tomograms. The information contained in the low amplitude and high frequency flickering is lost first. The flickering information is also lost if long integration times are used. From the wind simulations we conclude that a line optical depth greater than 10 is needed to obtain single peaked wind line profiles.\\
As incoming work this code will be used to generate synthetic tomograms with flickering, which will be compared to observed flickering tomograms of V3885 Sgr, aiming to locate the sites of flickering production and constrain the flickering parameters in this system.

\begin{acknowledgements}
F.M.A.R is grateful from support from FAPESP fellowship 01/07078-8. MD acknowledges the support by CNPq under grant \#301029.
\end{acknowledgements}

\end{article}
\end{document}